\documentclass[twocolumn,aps,pra,amsfonts,amssymb]{revtex4-1}
\usepackage{graphicx}
\usepackage{amsmath}
\usepackage{bm}

\newcounter{gg}

\newcommand{\nuab}{\bar{\nu}_a}

\newcommand{\nuzb}{\bar{\nu}_z}

\newcommand{\fcb}{\bar{f}_c}

\begin{document}

\title{New Measurement of the Electron Magnetic Moment and the Fine Structure Constant}

\author{D.\ Hanneke}
\affiliation{Department of Physics, Harvard University, Cambridge, MA 02138}

\author{S.\ Fogwell}
\affiliation{Department of Physics, Harvard University, Cambridge, MA 02138}

\author{G.\ Gabrielse}
\email[Email: ]{gabrielse@physics.harvard.edu} \affiliation{Department of Physics, Harvard University, Cambridge, MA
02138}

\date{Phys.\ Rev.\ Lett. {\bf 100}, 120801 (2008)}

\begin{abstract}     
A measurement using a one-electron quantum cyclotron gives the electron magnetic moment in Bohr magnetons, $g/2 = 1.001
\, 159 \, 652 \, 180 \, 73 \, (28) \, [0.28\, \rm{ppt}]$, with an uncertainty 2.7 and 15 times smaller than for
previous measurements in 2006 and 1987. The electron is used as a magnetometer to allow lineshape statistics to
accumulate, and its spontaneous emission rate determines the correction for its interaction with a cylindrical trap
cavity. The new measurement and QED theory determine the fine structure constant, with $\alpha^{-1}=137.035 \, 999 \,
084 \, (51) \, [0.37 \, \rm{ppb}]$, and an uncertainty 20 times smaller than for any independent determination of
$\alpha$.
\end{abstract}


\maketitle

\newcommand{\w}{3.5in}

\newcommand{\HistoryFigure}[1][\w]{
\begin{figure}[htbp!]
\centering
\includegraphics*[width=3.25in]{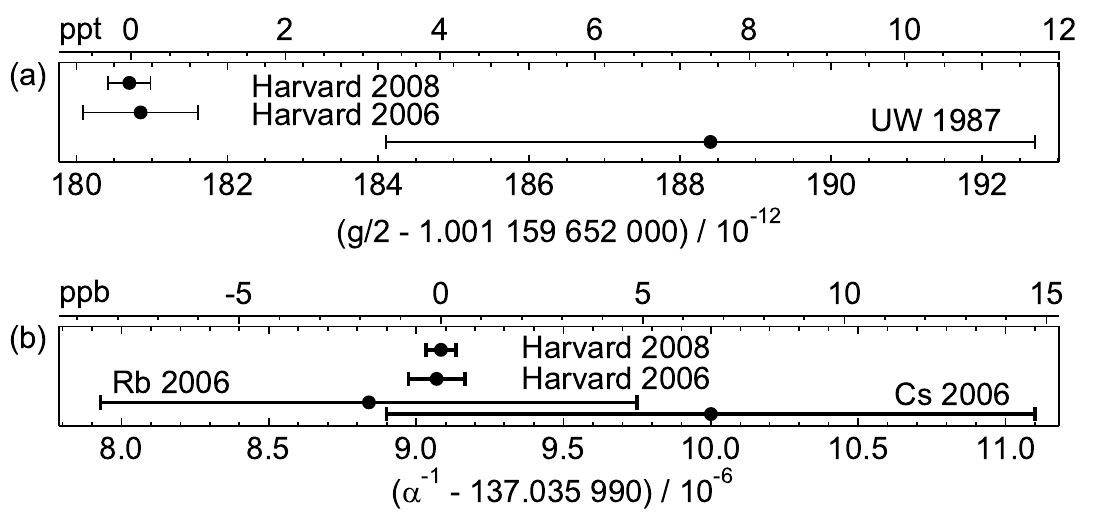}
\caption{Most accurate measurements of the electron $g/2$ (a), and most accurate determinations of $\alpha$ (b).}
\label{fig:History}
\end{figure}
}

\newcommand{\TrapAndLevelsFigure}[1][\w]{
\begin{figure}[htbp!]
\centering
\includegraphics*[width=3.25in]{TrapAndEnergyLevels}
\caption{Cylindrical Penning trap cavity used to confine a single electron and inhibit spontaneous emission (a), and
the cyclotron and spin levels of an electron confined within it (b).} \label{fig:TrapAndLevels}
\end{figure}
}

\newcommand{\TrapFigure}[1][\w]{
\begin{figure}[htbp!]
\centering
\includegraphics*[width=2.5in]{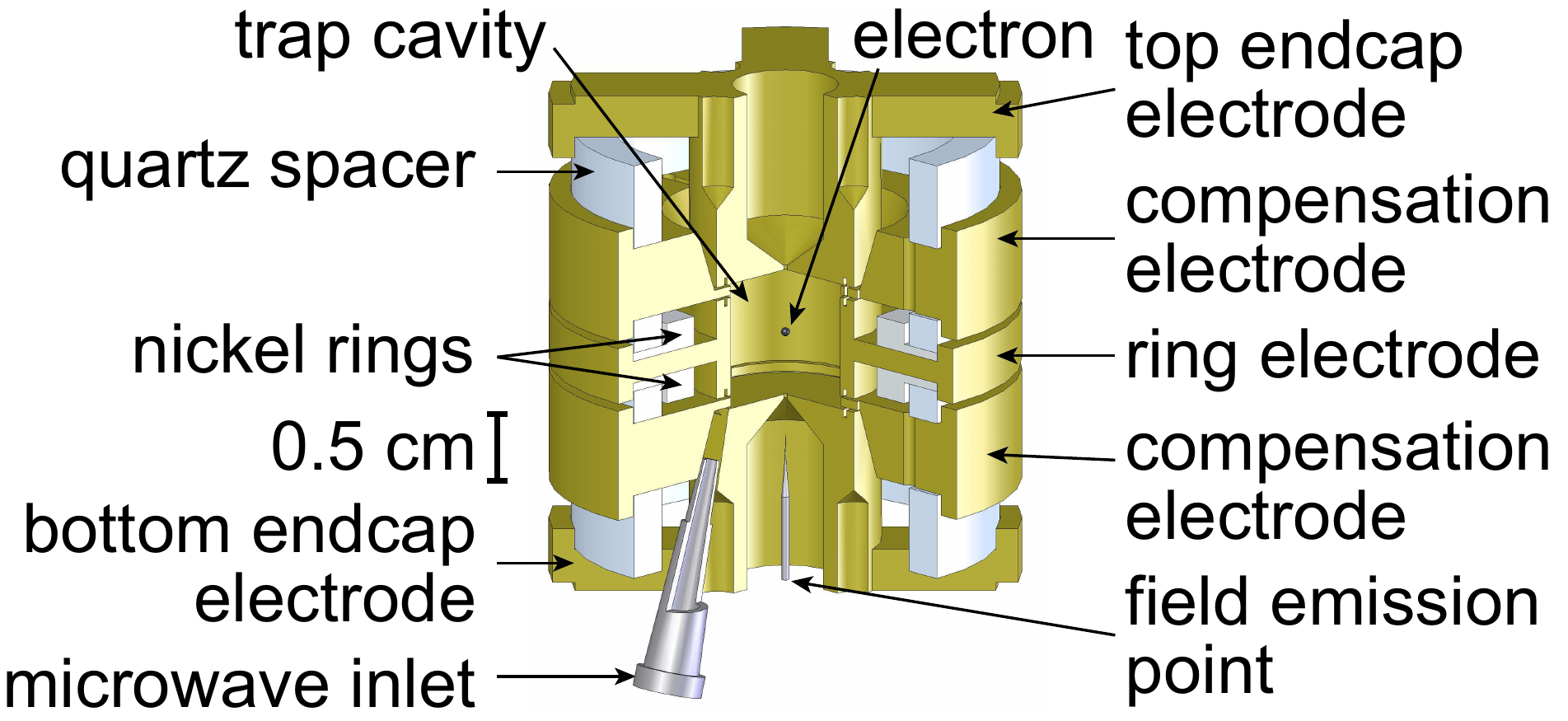}
\caption{Cylindrical Penning trap cavity used to confine a single electron and inhibit spontaneous emission.} \label{fig:Trap}
\end{figure}
}

\newcommand{\EnergyLevelsFigure}[1][\w]{
\begin{figure}[htbp!]
\centering
\includegraphics*[width=1.5in]{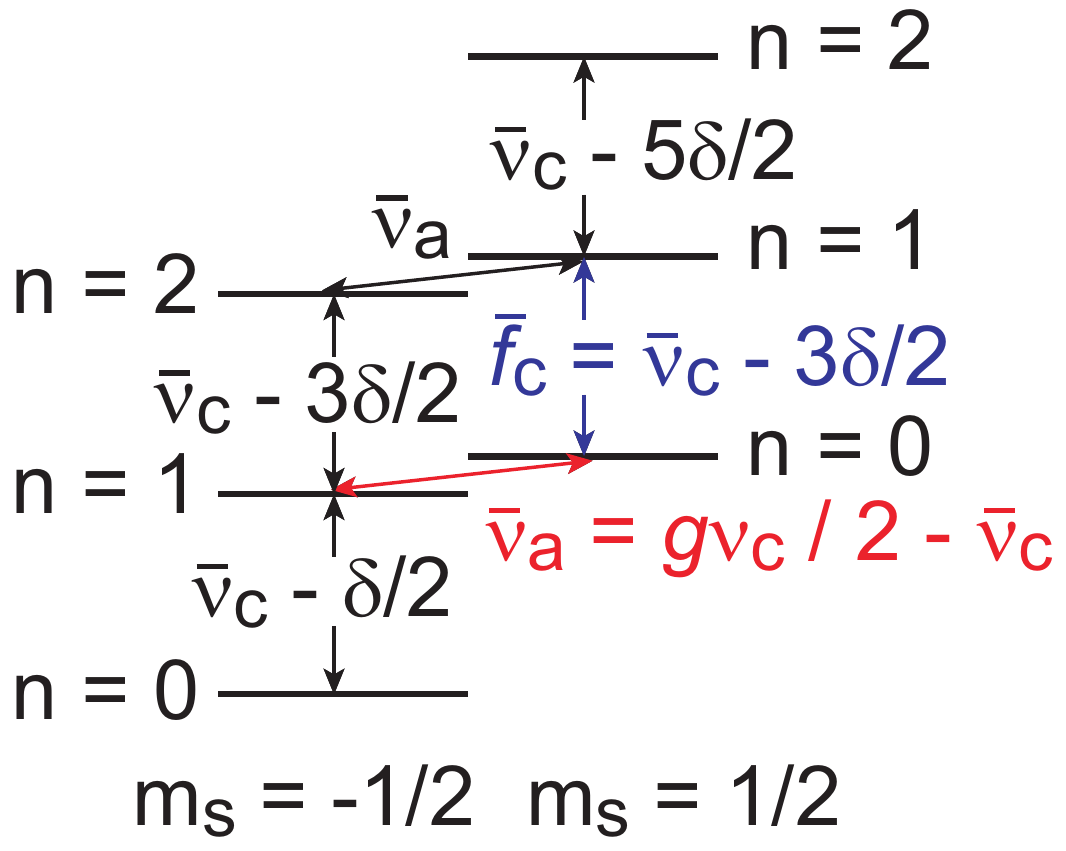}
\caption{Electron's lowest cyclotron and spin levels.} \label{fig:EnergyLevels}
\end{figure}
}

\newcommand{\LineShapesFigure}[1][\w]{
\begin{figure}[htbp!]
\centering
\includegraphics*[width=3.25in]{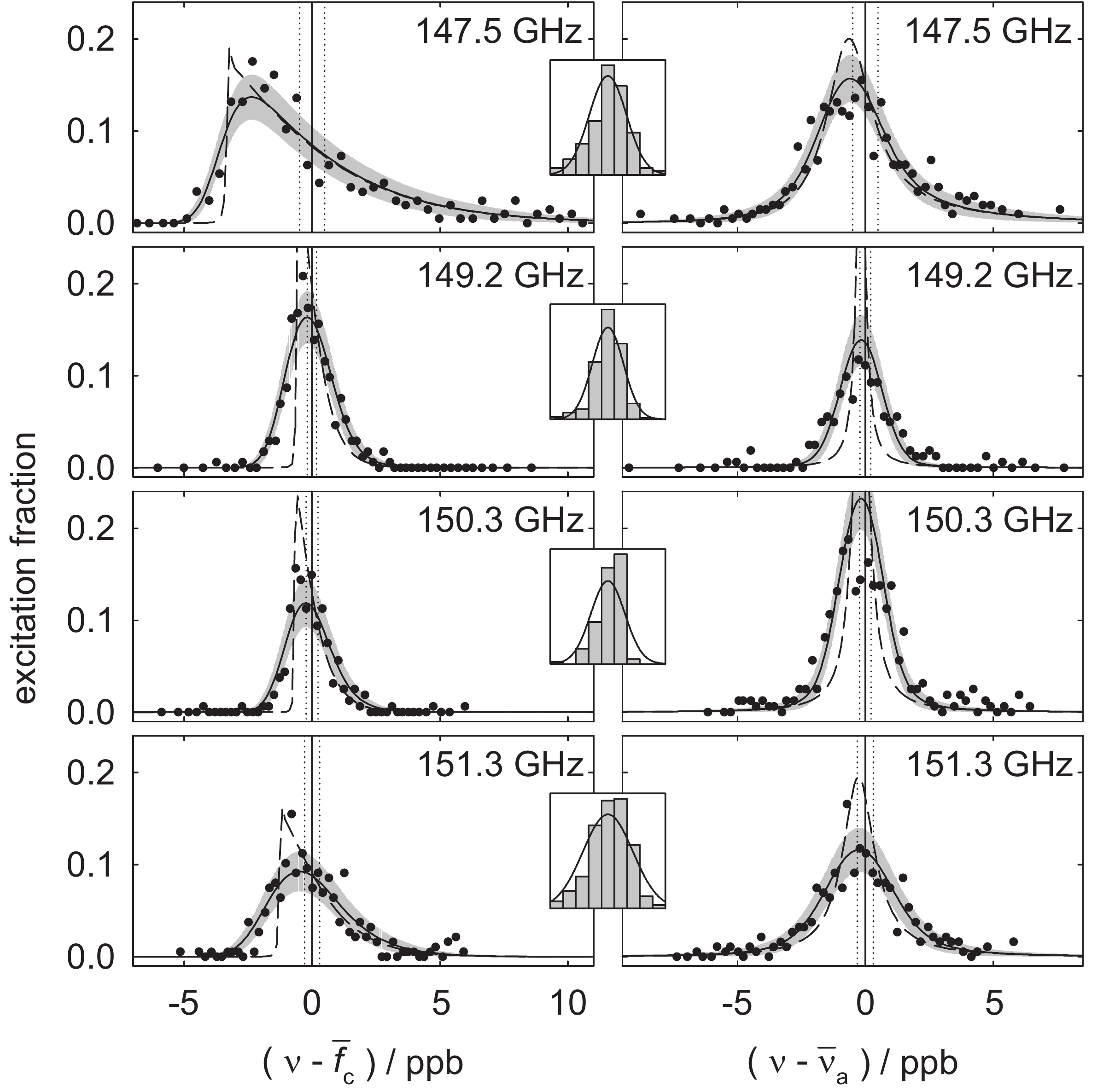}
\caption{Quantum-jump spectroscopy lineshapes for cyclotron (left) and anomaly (right) transitions, with maximum
likelihood fits to broadened lineshape models (solid), and inset resolution functions. Vertical lines show the
1-$\sigma$ uncertainties for extracted resonance frequencies. Corresponding un-broadened lineshapes are dashed. Gray
bands indicate 68\% confidence limits for distributions about broadened fits.  } \label{fig:CyclotronAndAnomalyLines}
\end{figure}
}

\newcommand{\CavityShiftsFigure}[1][\w]{
\begin{figure}[htbp!]
\centering
\includegraphics*[width=3.25in]{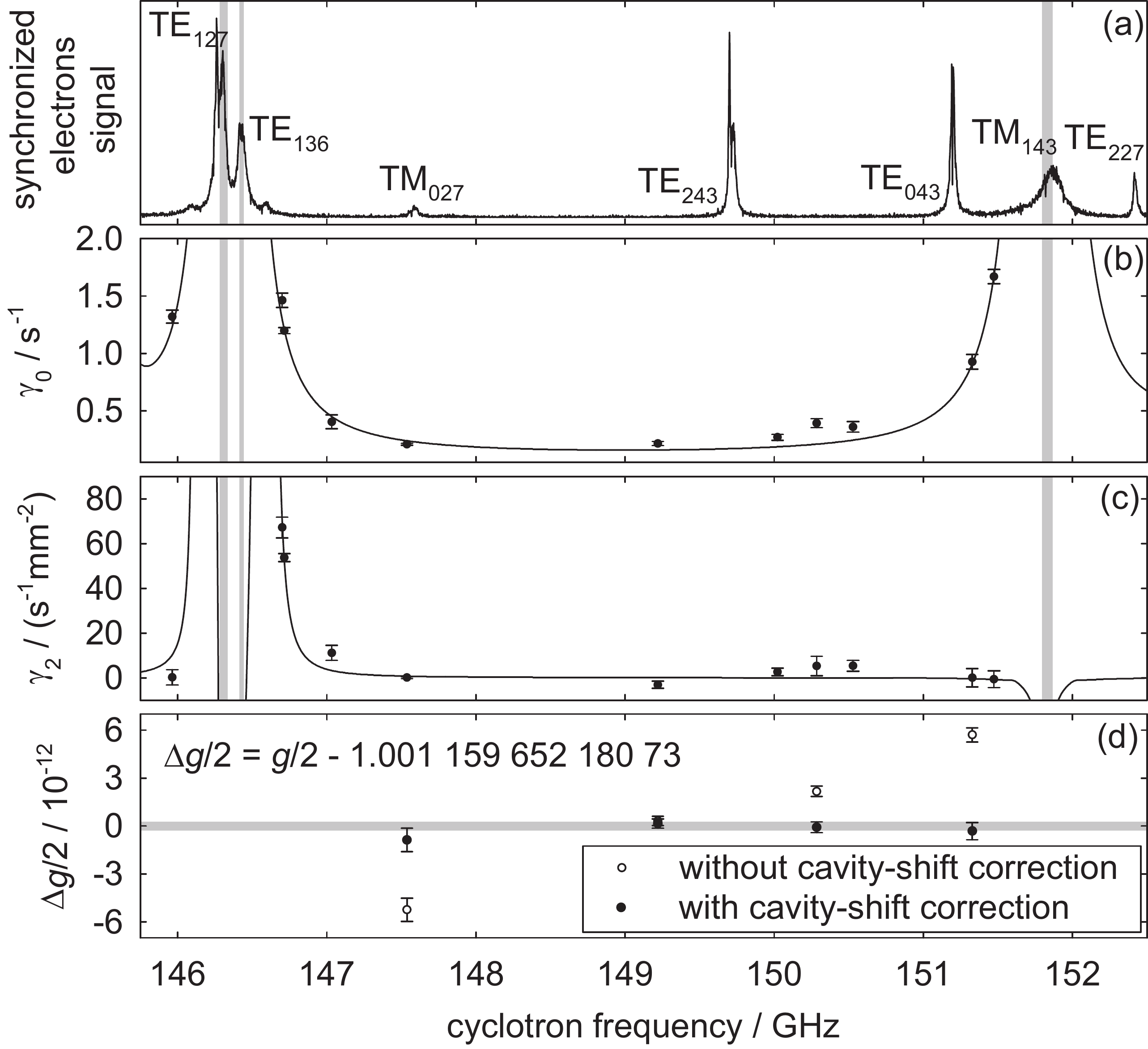}
\caption{Modes of the trap cavity are observed with synchronized electrons (a) \cite{HarvardMagneticMoment2006},  as well
as with a single electron damping rate $\gamma_0$ (b) and its amplitude dependence $\gamma_2$ (c). Offset of $g/2$ from
our result in Eq.~\ref{eq:g} without (open circle) and with (points) cavity-shift corrections, with an uncertainty band
for the average (d).} \label{fig:CavityShifts}
\end{figure}
}

The electron magnetic moment $\boldsymbol{\mu}$ is one of the few measurable properties of one of the simplest of
elementary particles -- revealing its interaction with the fluctuating QED vacuum, and probing for size or composite
structure not yet detected.   What can be accurately measured is $g/2$, the magnitude of $\boldsymbol{\mu}$ scaled by
the Bohr magneton, $\mu_B=e\hbar/(2m)$. For an eigenstate of spin ${\bf S}$,
\begin{equation}
\boldsymbol{\mu}=- \frac{g}{2} \, \mu_B \, \frac{\mathbf{S}}{\hbar/2},
\end{equation}
with $g/2=1$ for a point electron in a renormalizable Dirac description. QED predicts that vacuum fluctuations and
polarization slightly increase this value. Physics beyond the standard model of particle physics could make $g/2$
deviate from the Dirac/QED prediction (as internal quark-gluon substructure does for a proton).

The 1987 measurement that provided the accepted $g/2$ for nearly 20 years \cite{DehmeltMagneticMoment} was superceded
in 2006 by a measurement that used a one-electron quantum cyclotron \cite{HarvardMagneticMoment2006}. Key elements were
quantum jump spectroscopy and quantum non-demolition (QND) measurements of the lowest cyclotron and spin levels
\cite{QuantumCyclotron}, a cylindrical Penning trap cavity \cite{CylindricalPenningTrap} (Fig.~\ref{fig:Trap}),
inhibited spontaneous emission \cite{InhibitionLetter}, and a one-particle self-excited oscillator (SEO)
\cite{SelfExcitedOscillator}. This Letter reports an improved measurement that has a 2.7 and 15 times lower uncertainty
than the 2006 and 1987 measurements, respectively, and confirms a 1.8 standard deviation shift of the 1987 value
(Fig.~\ref{fig:History}a).  The interaction of the electron and its surrounding trap cavity is probed by measuring
$g/2$ and the electron's spontaneous emission rate as a function of magnetic field, thereby determining the corrections
needed for good agreement between measurements at different fields. The electron is also used as its own magnetometer
to accumulate quantum-jump lineshape statistics over days, making it possible to compare methods for extracting the
resonance frequencies.

\HistoryFigure

The new measurement and recently updated QED theory \cite{RevisedC8} determine $\alpha$ with an uncertainty 20 times smaller than does any
independent method (Fig.~\ref{fig:History}b). The uncertainty in $\alpha$ is now limited a bit more by the need for a higher-order QED calculation
(underway \cite{RevisedC8}) than by the measurement uncertainty in $g/2$.   The accuracy of the new $g$ sets the stage for an improved CPT test with
leptons. It also will allow an improved test of QED, and will be part of the discovery of low-mass dark-matter particles or the elimination of this
possibility \cite{LightDarkMatterArchive}, when a better independent measurement of $\alpha$ becomes available.

\TrapFigure

\EnergyLevelsFigure

Fig.~\ref{fig:EnergyLevels} represents the lowest cyclotron and spin energy levels for an electron weakly confined in a vertical magnetic field
$B{\bf \hat{z}}$ and an electrostatic quadrupole potential.  The latter is produced by biasing the trap electrodes of Fig.~\ref{fig:Trap}. The
measured cyclotron frequency $\fcb \approx 149$ GHz (blue in Fig.~\ref{fig:EnergyLevels}) and the measured anomaly frequency $\nuab \approx 173$ MHz
(red in Fig.~\ref{fig:EnergyLevels}) mostly determine $g/2$ \cite{HarvardMagneticMoment2006}
\begin{equation}
\frac{g}{2} \simeq 1 + \frac{\nuab - \nuzb^2/(2 \fcb)}{\fcb + 3\delta/2 + \nuzb^2/(2 \fcb)} + \frac{\Delta g_{cav}}{2}, \label{eq:Experimentalg}
\end{equation}
with only small adjustments for the measured axial frequency $\nuzb\approx 200$ MHz, the relativistic shift $\delta /\nu_c \equiv h\nu_c/(mc^2)
\approx 10^{-9}$, and the cavity shift $\Delta g_{cav}/2$.  The latter is the fractional shift of the cyclotron frequency caused by the interaction
with radiation modes of the trap cavity. The Brown-Gabrielse invariance theorem \cite{InvarianceTheorem} has been used to eliminate the effect of
both quadratic distortions to the electrostatic potential, and misalignments of the trap electrode axis with $\mathbf{B}$. Small terms of higher
order in $\nuzb /\fcb$ are neglected.

Quantum jump spectroscopy determines $\fcb$ and $\nuab$.  For each of many trials the system is prepared in the spin-up ground state,
$\left|n=0,m_s=1/2\right\rangle$, after which the preparation drives and detection amplifier are turned off for $1$ s. Either a cyclotron drive at a
frequency near to $\fcb$, or an anomaly drive at frequency near $\nuab$, is then applied for $2$ s.  The amplifier and a feedback system are turned
on to provide QND detection of either a one-quantum cyclotron excitation or a spin flip. Cavity-inhibited spontaneous emission makes the cyclotron
excitation persist long enough to allow such detection. Fig.~\ref{fig:CyclotronAndAnomalyLines} shows the fraction of the trials for which
excitations were detected.

\LineShapesFigure

The cyclotron drive is microwave radiation injected into the trap cavity through a cold attenuator to keep black body photons from entering the trap.
The anomaly drive is an oscillatory potential applied to electrodes at frequencies near $\nuab$ to drive off-resonant axial motion through the
magnetic bottle gradient from two nickel rings (Fig.~\ref{fig:Trap}). The electron, radially distributed as a cyclotron eigenstate, sees an
oscillating magnetic field perpendicular to $\mathbf{B}$ as needed to flip its spin, with a gradient that allows a simultaneous cyclotron transition
\cite{Palmer}. To ensure that the electron samples the same magnetic variations while $\nuab$ and $\fcb$ transitions are driven, both drives are kept
on with one detuned slightly so that only the other causes transitions. Low drive strengths keep transition probabilities below 20\% to avoid
saturation effects.

QND detection of one-quantum changes in the cyclotron and spin energies takes place because the magnetic bottle shifts the oscillation frequency of
the self-excited axial oscillation as $\Delta \nuzb \approx 4~(n + m_s)$ Hz. After a cyclotron excitation, cavity-inhibited spontaneous emission
provides the time needed to turn on the electronic amplification and feedback, so the SEO can reach an oscillation amplitude at which the shift can be
detected \cite{SelfExcitedOscillator}. An anomaly transition is followed by a spontaneous decay to the spin-down ground state,
$\left|n=0,m_s=-1/2\right\rangle$, and the QND detection reveals the lowered spin energy.

The expected lineshapes arise from the thermal axial motion of the electron through the magnetic bottle gradient. The axial motion is cooled by a
resonant circuit in about 0.2 s to as low as $T_z = 230$ mK (from 5 K) when the detection amplifier is off. For the cyclotron motion these
fluctuations are slow enough that the lineshape is essentially a Boltzmann distribution with a width proportional to $T_z$ \cite{BrownLineShape}. For
the anomaly resonance, the fluctuations are effectively more rapid, leading to a resonance shifted in proportion to $T_z$.

The weighted averages of $\nuab$ and $\fcb$ from the lineshapes (indicated by the abscissa origins in Fig.~\ref{fig:CyclotronAndAnomalyLines})
determine $g/2$ via Eq.~\ref{eq:Experimentalg}. With saturation effects avoided, these pertain to the magnetic field averaged over the thermal
motion. It is crucial that any additional fluctuations in $B$ that are symmetric about a central value will broaden such lineshapes without changing
the mean frequency.

To test this weighted mean method we compare maximum likelihood fits to lineshape models
(Fig.~\ref{fig:CyclotronAndAnomalyLines}).  The data fit well to a convolution (solid curve) of a Gaussian resolution
function (solid inset curve) and a thermal-axial-motion lineshape \cite{BrownLineShape} (dashed curve). The broadening
may arise from vibrations of the trap and electron through the slightly inhomogeneous field of the external solenoid,
or from fluctuations of the solenoid field itself. Because we have not yet identified its source we add a ``lineshape''
uncertainty based upon the discrepancy (beyond statistical uncertainty) between the $g/2$ values from the mean and fit
for the four measurements. To be cautious we take the minimum discrepancy as a correlated uncertainty, and then add the
rest as an uncorrelated uncertainty. An additional probe of the broadening comes from slowly increasing the microwave
frequency until a one-quantum cyclotron excitation is seen. The distribution of excitations in the inset histograms in
Fig.~\ref{fig:CyclotronAndAnomalyLines} are consistent with the Gaussian resolution functions determined from the fits.

Drifts of B are reduced below $10^{-9}$/hr by regulating five He and N$_2$ pressures in the solenoid and experiment
cryostats, and the surrounding air temperature \cite{HarvardMagneticMoment2006}.  Remaining slow drift is corrected using
the average of the described histograms taken once every three hours. Unlike the one-night-at-a-time analysis used in
2006, all data taken in four narrow ranges of $B$ values (Table \ref{table:guncertainties}) are combined, giving a
lineshape  signal-to-noise that allows the systematic investigation of lineshape uncertainty.

Better measurement and understanding of the electron-cavity interaction removes cavity shifts as a major uncertainty. Cavity shifts are the downside
of the cavity-inhibited spontaneous emission which usefully narrows resonance lines and gives the averaging time we need to turn on the SEO and
determine the cyclotron state.  The shifts arise when the cyclotron oscillator has its frequency pulled by the coupling to nearby radiation modes of
the cavity. The cylindrical trap cavity was invented \cite{CylindricalPenningTrap} and developed \cite{CylindricalPenningTrapDemonstrated} to deliberately modify the
density of states of the free space radiation modes in a controllable and understandable way (though not enough to require modified QED calculations
\cite{BrownApparatusDependentg}). Radiation mode frequencies must still be measured to determine the effective dimensions of a right-circular cylindrical cavity
which has been imperfectly machined, which has been slit (so sections of the cavity can be separately biased trap electrodes), and whose dimensions
change as the electrodes cool from 300 to 0.1 K.

\CavityShiftsFigure

To the synchronized-electrons method used earlier we add a new method -- using the electron itself to determine the cavity-electron interaction. The
measured spontaneous emission rate for its cyclotron motion, $\gamma = \gamma_0 + \gamma_2 A^2$, depends upon the amplitude $A$ of the axial
oscillation through the standing waves of cavity radiation modes.  $A$ is varied by adjusting the SEO \cite{SelfExcitedOscillator} and measured by
fitting to a cyclotron quantum-jump lineshape \cite{SelfExcitedOscillator,BrownLineShape}. Fits of $\gamma_0$ and $\gamma_2$
(Fig.~\ref{fig:CavityShifts}b-c) to a renormalized calculation of the coupling of the electron and cavity \cite{RenormalizedModesPRL} determine the
frequencies (with uncertainties represented by the vertical gray bands in Fig.~\ref{fig:CavityShifts}a-c) and $Q$ values of the nearest cavity modes,
and the cavity-shift corrections for $g/2$ (Table \ref{table:guncertainties}). (Subtleties in applying this calculation to measurements will be
reported separately.) Substantially different cavity-shift corrections bring the four $g/2$ measurements into good agreement
(Fig.~\ref{fig:CavityShifts}d).

\begin{table}
   \centering  
     \begin{tabular}{lr@{\,(}lrrr}
               \hline\hline 
               $\fcb$    & \multicolumn{2}{r}{147.5 GHz} & 149.2 GHz& 150.3 GHz& 151.3 GHz\\
               \hline\hline
               $g/2$ raw &-5.24&0.39)& 0.31\,(0.17) & 2.17\,(0.17)& 5.70\,(0.24) \\
               \hline

                   Cav.\ shift    & 4.36&0.13) & -0.16\,(0.06) & -2.25\,(0.07) & -6.02\,(0.28) \\

                   Lineshape & \multicolumn{5}{r}{}\\
                   \multicolumn{2}{l@{\,(}}{~~correlated}        & 0.24) & (0.24) & (0.24) & (0.24) \\
                   \multicolumn{2}{l@{\,(}}{~~uncorrelated}      & 0.56) & (0.00) & (0.15) & (0.30) \\
\hline
                   $g/2$ & -0.88&0.73) & 0.15\,(0.30) & -0.08\,(0.34) & -0.32\,(0.53) \\
               \hline\hline
           \end{tabular}
       \caption{Measurements and shifts with uncertainties, all multiplied by $10^{12}$.  The cavity-shifted ``$g/2$ raw'' and
       corrected ``$g/2$'' are offset from our result in
Eq.~\ref{eq:g}.}\label{table:guncertainties}
\end{table}

The measured values, shifts, and uncertainties for the four separate measurements of $g/2$ are in Table~\ref{table:guncertainties}.  The
uncertainties are lower for measurements with smaller cavity shifts and smaller linewidths, as might be expected.  Uncertainties for variations of
the power of the $\nuab$ and $\fcb$ drives are estimated to be too small to show up in the table. A weighted average of the four measurements, with
uncorrelated and correlated errors combined appropriately, gives the electron magnetic moment in Bohr magnetons,
\begin{equation}
g/2 = 1.001 \, 159 \, 652 \, 180 \, 73 \, (28)~~~~~[0.28~\rm{ppt}]. \label{eq:g}
\end{equation}
The uncertainty is 2.7 and 15 times smaller than the 2006 and 1987 measurements, and $2300$ times smaller than has been
achieved for the heavier muon lepton \cite{gMuon2006}.

The new measurement determines the fine structure constant, $\alpha = e^2/(4\pi\epsilon_0 \hbar c)$, the fundamental
measure of the strength of the electromagnetic interaction in the low energy limit, that is also a crucial ingredient
of our system of fundamental constants \cite{CODATA2002}. The standard model relates $g$ and $\alpha$ by
\begin{eqnarray}
\frac{g}{2}= &1& + \, C_2\left(\frac{\alpha}{\pi}\right) +C_4\left(\frac{\alpha}{\pi}\right)^2
+C_6\left(\frac{\alpha}{\pi}\right)^3
+C_8\left(\frac{\alpha}{\pi}\right)^4 \nonumber\\
&+&C_{10}\left(\frac{\alpha}{\pi}\right)^5 + ... + a_{\mu\tau} + a_{\rm{hadronic}} + a_{\rm{weak}},
\label{eq:QedSeries}
\end{eqnarray}
with the asymptotic series and $a_{\mu\tau}$ coming from QED.  Very small hadronic and weak contributions are included,
along with the assumption that there is no significant modification from electron substructure or other physics beyond
the standard model. Calculations summarized in \cite{Alpha2006fixed} give exact $C_2$, $C_4$ and $C_6$, a numerical
value and uncertainty for $C_8$, and a small $a_{\mu\tau}$. The result is
\begin{eqnarray}
\alpha^{-1} &=& 137.035 \, 999 \, 084 \, (33)\,(39)~~[0.24~\rm{ppb}]\,[0.28~\rm{ppb}],\nonumber \\
&=& 137.035 \, 999 \, 084 \, (51)~~~~~~~~[0.37~\rm{ppb}]. \label{eq:AlphaValue}
\end{eqnarray}
The first line shows experimental (first) and theoretical (second) uncertainties that are nearly the same.  The total $0.37$ ppb uncertainty in
$\alpha$ is 20 times smaller than for the next most precise independent methods (Fig.~\ref{fig:History}b).  These so-called atom recoil methods
\cite{AlphaRb2006,Tanner2006} utilize measurements of transition frequencies and mass ratios, as well as either a Rb recoil velocity
(in an optical lattice) or a Cs recoil velocity (in an atom interferometer).

The theory uncertainty contribution to $\alpha$ is divided as $(12)$ and $(37)$ for $C_8$ and $C_{10}$.  It should
decrease when a calculation underway \cite{RevisedC8} replaces the crude estimate $C_{10}=0.0 \,(4.6)$
\cite{CODATA2002,Alpha2006fixed}.  The $\alpha^{-1}$ of Eq.~\ref{eq:AlphaValue} will then shift by $2\alpha^3 \pi^{-4}
C_{10}$, which is $8.0\, C_{10} \times 10^{-9}$.  A change $\Delta_8$ in the calculated $C_8=-1.9144\,(35)$ would add
$2\alpha^2\pi^{-3}\Delta_8$.

The new $g/2$ allows three additional applications if a way is found to measure $\alpha$ independently  at our
accuracy.  First, is a 20 times more stringent test of QED. Second, is a 20 times more sensitive probe for electron
size and substructure \cite{Alpha2006fixed}. Third, is a 20 times more sensitive search for a dark matter particle of
low mass \cite{LightDarkMatterArchive}.

Items that warrant further study could lead to a future measurement of $g/2$ to higher precision.  First is the broadening of the expected lineshapes
which limits the splitting of the resonance lines.  Second, the variation in axial temperatures in Fig.~\ref{fig:CyclotronAndAnomalyLines}, not
understood, increases the uncertainty contributed by the wider lineshapes.  Third, cavity sideband cooling could cool the axial motion to near its
quantum ground state for a more controlled measurement.  Fourth, a new apparatus should be much less sensitive to vibration and other variations in
the laboratory environment.

In conclusion, a new measurement of the electron $g/2$ is $15$ times more accurate than the 1987 measurement that
provided $g/2$ and $\alpha$ for nearly 20 years, and $2.7$ times more accurate than the 2006 measurement that
superseded it.  Achieving the reported electron $g/2$ uncertainty with a positron seems feasible, to make the most
stringent lepton CPT test.   With QED and the assumption of no new physics beyond the standard model of particle
physics, the new measurement determines $\alpha$ 20 times more accurately than any independent method.  The measured
$g/2$ is accurate enough to allow testing QED, probing for electron size, and searching for a low mass dark matter
particle if a more accurate independent measurement of $\alpha$ is realized.

More details will follow in a longer report \cite{HannekeToBePublished}. Thanks for help and comments to Y.\ Gurevich and B.\ Odom. This thesis work
of D. Hanneke \cite{ThesisHanneke} was supported by the NSF AMO program.


%
\end{document}